\documentclass[twoside,fleqn]{article}
\usepackage{espcrc2}
\usepackage{psfig}
\title{N* Masses from an Anisotropic Lattice QCD Action}
\author{F.X.~Lee
\address[GW]{Center for Nuclear Studies,
        George Washington University, Washington, DC 20052, USA}
\address[JL]{Jefferson Lab, 12000 Jefferson Avenue, Newport News, VA 23606, USA},
D.B.~Leinweber\address[CSSM]{Department of Physics and Mathematical
Physics and CSSM, Adelaide University, SA 5005, Australia},
L.~Zhou\addressmark[GW],
J. Zanotti\addressmark[CSSM],
S.~Choe\address{Hiroshima University, Higashi-Hiroshima 739-8526, Japan}}
\begin{document}
\begin{abstract}
We report N* masses in the spin-3/2 sector 
from a highly-improved anisotropic action.
States with both positive and
negative parity are isolated via a parity projection method.  The
extent to which spin projection is needed 
is examined. The gross features of
the splittings from the nucleon ground state show a trend consistent
with experimental results at the quark masses explored.  
\end{abstract}
\maketitle

\section{Introduction}
Lattice QCD plays an important role in understanding the N* spectrum.
One can systematically study the spectrum sector by sector, with the ability 
to dial the quark masses, and dissect the degrees of freedom.
The rich structure of the excited baryon spectrum, as tabulated by the 
particle data group~\cite{pdg00},
provides a fertile ground for exploring how the internal 
degrees of freedom in the nucleon are excited 
and how QCD works in a wider context.
One outstanding example is the parity splitting pattern
in the low-lying N* spectrum.
The splittings are a direct manifestation of spontaneous chiral symmetry breaking 
of QCD.  Without it, QCD predicts parity doubling in the baryon spectrum.
The experimental effort on excited baryons has intensified in recent years 
at JLab and other accelerators, generating renewed debate on how well these states 
are known. The star-rating system on the observed states 
is a reflection of the current situation.

Given that state-of-the-art lattice QCD simulations have produced a ground-state
spectrum that is very close to the observed values~\cite{cppacs00},
it is important to extend the success beyond the ground states.
There exist already a number of lattice studies of the
N* spectrum~\cite{derek95,lee99,lee00,sas00,dgr00,dgr01},
focusing mostly on the spin-1/2 sector.
All established a clear splitting from the ground state.
In this study, we explore the possibility of calculating
the excited baryon states in the spin-3/2 sector with isospin 1/2.

\section{Method}
We consider the following interpolating field with the quantum numbers 
$I(J^P)={1\over 2}\left({3\over 2}^+\right)$~\cite{chung82,derek90},
\begin{equation}
\chi_\mu = \epsilon^{abc}\left( u_a^{T}C\gamma_5\gamma_\rho d_b\right) 
\left(g^{\mu\rho}-{1\over 4}\gamma^\mu\gamma^\rho\right) \gamma_5 u_c.
\end{equation}
Despite having an explicit parity by construction, the interpolating
field couples to both positive and negative parity states.
A parity projection is needed to separate the two.
In the large Euclidean time limit, 
the correlator with Dirichlet boundary condition in the time direction
and zero spatial momentum becomes
\begin{equation}
G_{\mu\nu}(t) = \sum_{\bf x} <0|\chi_\mu(x)\,\overline\chi_\nu(0)|0>
\end{equation}
\begin{equation}
= f_{\mu\nu}\left[\lambda_+^2 {\gamma_4 +1 \over 2} e^{- M_+ \, t}
+ \lambda_-^2 {-\gamma_4 +1 \over 2} e^{- M_- \, t} \right]
\end{equation}
where $f_{\mu\nu}$ is a function common to both terms.
The relative sign in front of $\gamma_4$ provides the solution: by taking the
trace of $G_{\mu\nu}(t)$ with $(1\pm\gamma_4)/4$, one can isolate 
$M_+$ and $M_-$, respectively. 

It is well-known that a spin-3/2 interpolating field couples to both 
spin-3/2 and spin-1/2 states. 
A spin projection can be used to isolate the individual contributions 
in the correlation function $G_{\mu \nu}$.
Using the spin-3/2 projection operator~\cite{derek92},
\begin{equation}
P_{\mu \nu }^{3/2} = g_{\mu \nu } - \frac{1}{3} \gamma_\mu
\gamma_\nu  - \frac{1}{3 p^2} ( \gamma \cdot p \gamma_\mu p_\nu +
p_\mu \gamma_\nu \gamma \cdot p )
\end{equation}
the spin-3/2 part can be projected out by 
\begin{equation}
G_{\mu \nu}^{3/2}(t) = \sum \limits_{\lambda =1}^4 G_{\mu \lambda}(t)
P_{\lambda \nu}^{3/2},
\end{equation}
while the spin-1/2 part by 
\begin{equation}
G_{\mu \nu}^{1/2}(t) = \sum \limits_{\lambda =1}^4 G_{\mu \lambda}(t)
(1-P_{\lambda \nu}^{3/2}).
\end{equation}
Obviously, they satisfy the relation
\begin{equation}
G_{\mu \nu}(t) = G_{\mu \nu}^{1/2}(t)+G_{\mu \nu}^{3/2}(t).
\label{sum}
\end{equation}

The anisotropic gauge action of~\cite{mor97}, and
the anisotropic D234 quark action of~\cite{alf98} are used.
Both have tadpole-improved tree-level coefficients.
A $10^3\times 30$ lattice with $a_s\approx 0.24$ fm  
and anisotropy $\xi=a_s/a_t=3$ is used.
In all, 100 configurations are analyzed.  On each configuration 
9 quark propagators are computed using a multi-mass solver, 
with quark masses ranging from approximately
780 to 90 MeV. The corresponding mass ratio $\pi/\rho$ is from 0.95 to 0.65.
A gauge-invariant gaussian-smeared source is used.
The source is located at $(x,y,z,t)=(2,2,2,2)$.

\begin{figure}
\centerline{\psfig{file=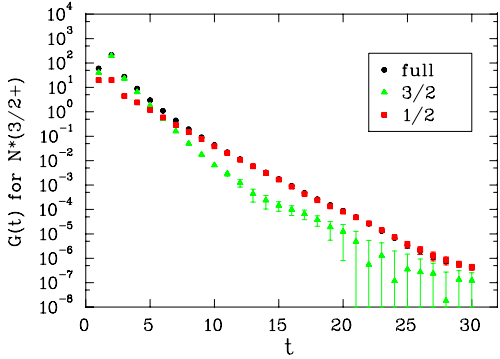,width=6.5cm,angle=90}}
\vspace*{-0.5cm}
\caption{The various correlation functions 
for the positive-parity nucleon states at the smallest quark mass.}
\label{corr_n3tr_pos}
\end{figure}

Figure~\ref{corr_n3tr_pos} presents results for the
correlation function in the positive-parity nucleon channel at the
smallest quark mass considered.  Spin projection reveals two different
exponentials from the spin-3/2 and spin-1/2 parts, with 
the spin-3/2 state being heavier than the spin-1/2 one (a steeper fall-off), 
in agreement with the ordering in experiment.  
The expected relation in Eq.~(\ref{sum}) is indeed satisfied numerically, 
providing a non-trivial check of the calculation.
A further check of the calculation is provided by the fact that 
the mass extracted from $G_{\mu \nu}^{1/2}(t)$ is degenerate with 
that from the conventional $G(t)$ for the nucleon ground state 
using the interpolating field 
$\chi = \epsilon^{abc}\left( u_a^{T}C\gamma_5 d_b\right)  u_c$.
One would get a false signal
from the dominant spin-1/2 state without spin projection.  The large
error bars indicate sensitive cancellations in the projection
procedure.  

\begin{figure}
\centerline{\psfig{file=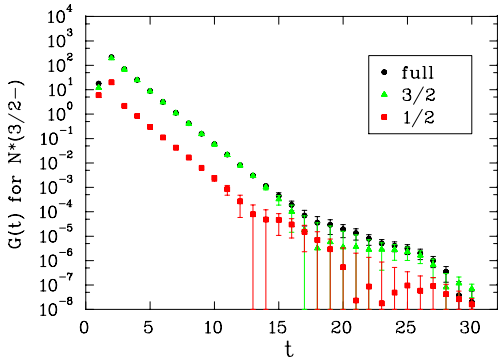,width=6.5cm,angle=90}}
\vspace*{-0.5cm}
\caption{Similar to Figure~\ref{corr_n3tr_pos}, but for the
negative-parity nucleon states.}
\label{corr_n3tr_neg}
\end{figure}

Figure~\ref{corr_n3tr_neg} shows the similar plot in the
negative-parity nucleon channel. Here the relation in Eq.~(\ref{sum})
is also satisfied, even though the signal is dominated by the $3/2-$ state.
The results also show a similar mass for the $1/2-$ state and the $3/2-$ state,
in accord with the experimental states of $N^*(1535){1\over 2}^-$ and 
$N^*(1520){3\over 2}^-$ which are close to each other. 

\begin{figure}
\centerline{\psfig{file=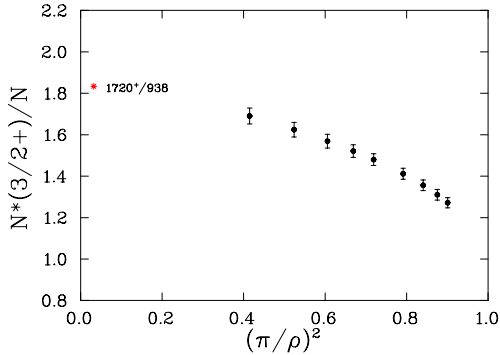,width=6.5cm,angle=90}}
\vspace*{-0.5cm}
\caption{
Mass ratio $N^*(3/2+)/N(1/2+)$ as a function of the mass ratio $(\pi/\rho)^2$. 
The experimental point is indicated for reference.}
\label{Ratio3_pos}
\end{figure}

\begin{figure}
\centerline{\psfig{file=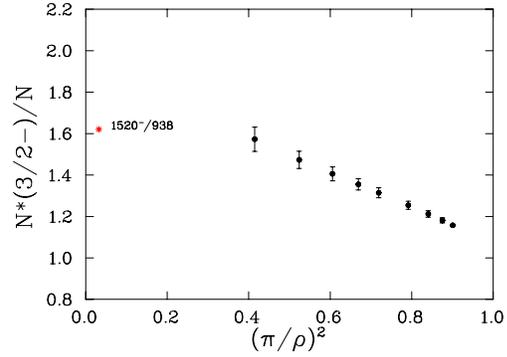,width=6.5cm,angle=90}}
\vspace*{-0.5cm}
\caption{Similar to Figure~\ref{Ratio3_pos}, but for the ratio 
$N^*(3/2-)/N(1/2+)$.}
\label{Ratio3_neg}
\end{figure}

Figure~\ref{Ratio3_pos} presents results for the mass
ratio extracted from the correlation functions for the $N^*(3/2^+)$ 
state to the nucleon ground state as a function of the mass ratio $(\pi/\rho)^2$.
Mass ratios have minimal dependence on the uncertainties in
determining the scale and the quark masses, so that a more reliable
comparison with experiment can be made.  
Figure~\ref{Ratio3_neg} shows the similar plot for the $N^*(3/2-)$ state.
These ratios appear headed in
the right direction compared to experiment.

\section{Conclusion} 
We have obtained clear signals for spin-3/2 N* states
on an anisotropic lattice with smeared operators and 100
configurations.  States of both negative- and positive-parity are
isolated with a parity projection technique.  The need for spin
projection is further demonstrated.  The results for the
pattern of the splittings appears consistent with experiment,
but more study is needed to address systematic errors.

\section{Acknowledgment}
This work is supported in part by U.S. Department of Energy 
under grant DE-FG02-95ER40907,
the Australian Research Council,
and is part of the effort by the Lattice Hadron Physics Collaboration~\cite{lhpc}.
The computing resources at NERSC and JLab have been used.


\end{document}